\documentclass[british,a4paper]{article}

\usepackage{babel,amsmath,amssymb}
\usepackage{bm}
\usepackage[sans]{dsfont}
\usepackage[mathscr]{euscript}
\usepackage{times}
\usepackage{cite}
\newcommand{\Tr}{\operatorname{Tr}}
\newcommand{\rmd}{\mathrm{d}}
\newcommand{\rme}{\mathrm{e}}
\newcommand{\rmi}{\mathrm{i}}
\newcommand{\RE}{\mathrm{\,Re\,}}
\newcommand{\IM}{\mathrm{\,Im\,}}
\newcommand{\Ebb}{\mathbb{E}}
\newcommand{\Rbb}{\mathbb{R}}
\newcommand{\Cbb}{\mathbb{C}}
\newcommand{\Zbb}{\mathbb{Z}}

\newcommand{\om}{{\omega_{\rm m}}}

\newcommand{\omq}{{\omega_{\rm m}^{\,2}}}
\newcommand{\Om}{{\Omega_{\rm m}}}
\newcommand{\Omq}{{\Omega_{\rm m}^{\,2}}}
\newcommand{\mgamq}{{\gamma_{\rm m}^{\,2}}}
\newcommand{\mgam}{{\gamma_{\rm m}}}
\newcommand{\rmm}{\mathrm{m}}
\newcommand{\openone}{\mathds{1}}

\newcommand{\norm}[1]{\left\Vert#1\right\Vert}
\newcommand{\abs}[1]{\left\vert#1\right\vert}

\newcommand{\Lcal}{\mathcal{L}}

 \newcommand{\Wcal}{\mathcal{W}}

\newcommand{\Hscr}{\mathscr{H}}

\title{Quantum stochastic equations for an opto-mechanical oscillator with radiation pressure interaction and non-Markovian effects}
\author{Alberto Barchielli\thanks{{} \ also: Istituto Nazionale di Fisica Nucleare (INFN), Sezione di Milano, and
Istituto Nazionale di Alta Matematica (INDAM-GNAMPA)} \\ Politecnico di Milano, Dipartimento di Matematica,
\\ piazza Leonardo da Vinci 32, 20133 Milano, Italy \\ (email: alberto.barchielli@polimi.it )
}

\begin{document}

\maketitle
\begin{abstract}
The quantum stochastic Schr\"odinger equation or Hudson-Parthasareathy (HP) equation is a powerful tool to construct unitary dilations of quantum dynamical semigroups and to develop the theory of measurements in continuous time via the construction of output fields. An important feature of such an equation is that it allows to treat not only absorption and emission of quanta, but also scattering processes, which however had very few applications in physical modelling. Moreover, recent developments have shown that also some non-Markovian dynamics can be generated by suitable choices of the state of the quantum noises involved in the HP-equation. This paper is devoted to an application involving these two features, non-Markovianity and scattering process. We consider a micro-mirror mounted on a vibrating structure and reflecting a laser beam, a process giving rise to a radiation-pressure force on the mirror.  We show that this process needs the scattering part of the HP-equation to be described. On the other side, non-Markovianity is introduced by the dissipation due to the interaction with some thermal environment which we represent by a phonon field, with a nearly arbitrary excitation spectrum, and by the introduction of phase noise in the laser beam. Finally, we study the full power spectrum of the reflected light and we show how the laser beam can be used as a temperature probe.

\end{abstract}

\markright{A. Barchielli: Quantum theory of an opto-mechanical oscillator}

\section{Introduction}\label{intro}
Quantum optomechanical systems represent an active field of research, very important both from the theoretical and experimental point of views, with applications in quantum optics and quantum information \cite{JacTWS99,GioV01,GMVT09,SN-P13,Chen13,asp14}. A great interest is due to the possibility of seeing quantum effects in a macroscopic mechanical resonator, say a mirror mounted on a vibrating structure and coupled to optical elements by radiation pressure. Typically, the theoretical description of such a kind of systems is based on the \emph{quantum Langevin equations} \cite{GarZ00,Car08}, a flexible approach allowing also for the introduction of non Markovian effects. Recently, some experimental evidence of non Markovian effects in an optomechanical system has been reported \cite{GroTru15}.

To get mathematically consistent quantum Langevin equations one has to use the \emph{quantum stochastic calculus} and the \emph{quantum stochastic Schr\"odinger equation}, or Hud\-son-Parthasarathy equation (HP-equation) \cite{HudP84,Parthas92}. In this mathematical context the quantum Langevin equations appear under the name of \emph{Evans-Hudson flows} or \emph{quantum flows} \cite{Parthas92,Fagnola99,LinW00}. In \cite{BarV15} a description of a dissipative mechanical oscillator has been obtained in terms of quantum stochastic differential equations; this description is fully consistent and valid at any temperature and it respects some symmetry requirements and physical constraints such as a weak form of equipartition at equilibrium and the translation invariance of the dissipative part of the dynamics. Non Markovian effects have been introduced by a suitable choice of the state of the \emph{quantum noises} appearing in the HP-equation. In Section \ref{LangTH} the quantum stochastic model of a dissipative mechanical oscillator is presented. An equation of Hudson-Parthasarathy type gives the unitary dynamics of the oscillator interacting with a Bose field (here representing the phonon field). The evolution equations for the system operators in the Heisenberg picture are the quantum Langevin equations and a suitable choice of the state of the field (based on a field analog of the $P$-representation in the case of discrete modes) allows for the introduction of thermal, non Markovian effects.

Usual quantum Langevin equations allow to describe absorption and emission of energy quanta by the main system, not scattering processes. As a matter of fact it seems that the existing literature takes into account the mirror/light radiation pressure interaction only if mediated by cavity modes; indeed the subject is often called \emph{cavity optmechanics} \cite{SN-P13,asp14}. This is due to the fact that some interesting physical phenomena as laser cooling appear when a cavity mode is involved \cite{GMVT09,SN-P13,Chen13,asp14,BarV15}, but also to the fact that a way to describe at a quantum level the direct scattering of laser light by a vibrating mirror is lacking. In this respect, another advantage of the HP-equation is that it allows also for the description of scattering processes \cite{BarL00,JG15}. The content of Section \ref{radpress} will be to introduce in the HP-equation, describing the mechanical oscillator, the radiation pressure interaction due to a laser directly illuminating the mirror. Finally, in Section \ref{hetdet} we show how to describe the heterodyne detection of the reflected light and we study the properties of the resulting power spectrum; this last step involves also the theory of measurements in continuous time \cite{Bar86,BarPer02,Bar06,BarG08b,BarCS11,BarG13}. In particular we show that, for a weak probe laser, the model produces explicit expressions for the spectrum due to elastic scattering of the photons and for the \emph{side-bands} due to \emph{Stokes and anti-Stokes scattering}.

We end this section by introducing the HP-equation. We give a short, heuristic presentation; a mathematically rigourous formulation of quantum stochastic calculus, HP-equations and related notions can be found in \cite{Parthas92,Bar06,BarCS11,FagW03,Fagnola99}. Firstly,
we introduce the formal fields $b_k(t)$, $b_k^\dagger(t)$, $t\in\Rbb$, $k=1\ldots d$, satisfying the canonical commutation rules (CCRs)
\begin{equation}\label{CCR}
\left[b_i(s),b_k^\dagger(t) \right]=\delta_{ik}\delta(t-s), \qquad \left[b_i(s),b_k(t) \right]=0.
\end{equation}
In this paper we consider only the representation of the CCRs \eqref{CCR} on the Fock space, the one characterized by the existence of the vacuum state. For quantum stochastic calculus involving non-Fock representations see, for instance, \cite{LinW86}. Let us introduce the Hilbert space $
L^2(\mathbb{R})\otimes\Cbb^d = L^2({\mathbb R}; \Cbb^d)$ (the one-particle space) and its symmetrized powers $L^2({\mathbb R}; \Cbb^d)^{\otimes_s n}$ (the $n$-particle space).
We denote by $\Gamma\equiv \Gamma\big(L^2({\mathbb R}; \Cbb^d)\big)$ the
\emph{symmetric Fock space} over $L^2({\mathbb R}; \Cbb^d)$, i.e.\ $\Gamma= \Cbb\oplus \sum_{n=1}^\infty L^2({\mathbb R}; \Cbb^d)^{\otimes_s n}$, and by $e(f)$,  $f
\in L^2({\mathbb R};\Cbb^d )$, \ the \emph{coherent vectors,} whose components
in the \ $0,1,\ldots,n,\ldots$ \ particle spaces are
\begin{equation}\label{expV}
e(f) :=\rme^{
-\frac{1}{2}\,\norm{f}^2}\left(1,f,(2!)^{-1/2}f\otimes f,\ldots,(n!)^{-1/2} f^{
\otimes n}, \ldots \right).
\end{equation}
Note that $e(0)$ represents the vacuum state and that
\[
\langle e(g)|e(f)\rangle = \exp\left\{ -\frac 1 2 \norm{f}^2 -\frac 1 2 \norm{g}^2 +\langle g|f\rangle\right\}.
\]
Let $\{z_k,\ k\geq 1\}$ be the canonical basis in $\Cbb^d$ and for any $f\in
L^2(\mathbb{R};\Cbb^d)$ let us set $f_k(t):= \langle z_k|f(t)\rangle_{\Cbb^d}$.
Then we have
$
b_k(t)\,e(f)= f_k(t)\, e(f)$.
By formally writing
\begin{equation}\label{Bb}
B_k(t)=\int_0^tb_k(s) \rmd s , \qquad B_k^\dagger(t)=\int_0^tb_k^\dagger(s) \rmd s,
\end{equation}
we get the \emph{annihilation} and \emph{creation processes}, families of
mutually adjoint operators, whose actions on the coherent vectors are given by
\[
B_k(t)\,e(f)= \int_0^t f_k(s) \,\rmd  s \, e(f)\,, \qquad
\langle e(g)| B_k^\dagger(t)e(f)\rangle = \int_0^t \overline{ g_k(s)}\, \rmd  s
\, \langle e(g)| e(f)\rangle.
\]
The overline denotes the complex conjugation. It is a property of the Fock spaces the fact that the action on the coherent
vectors uniquely determines a densely defined  linear operator. In terms of the integrated processes the CCRs \eqref{CCR} become
\begin{equation}\label{intCCR}
[B_k(t), B_l^\dagger(s)]=\delta_{kl}\min\{t,s\},\qquad [B_k^\dagger(t), B_l^\dagger(s)]=[B_k(t), B_l(s)]=0.
\end{equation}
We introduce also the \emph{gauge processes}
\begin{equation}\label{Lambda}
\Lambda_{kl}(t)=\int_0^t b_k^\dagger(s)b_l(s) \rmd s, \qquad
\langle e(g)| \Lambda_{kl}(t)e(f)\rangle = \int_0^t \overline{ g_k(s)}\, f_l(s)\rmd  s
\, \langle e(g)| e(f)\rangle.
\end{equation}
The operator $\Lambda_{kk}(t)$ turns out to be a number operator and it counts the quanta present in the field $k$ in the time interval $(0,t)$. Quantum stochastic calculus is an It\^o type calculus with respect to the integrators $\rmd t$, $\rmd B_k(t)$, $\rmd B_k^\dagger(t)$, $\rmd \Lambda_{kl}(t)$ satisfying the It\^o product rules
\begin{equation}\label{Itotab}\begin{split}
\rmd B_k(t)\rmd B_l^\dagger(t)=\delta_{kl}\rmd t, \qquad &\rmd B_i(t) \rmd \Lambda_{kl}(t)= \delta_{ik} \rmd B_l(t),\\
\rmd \Lambda_{kl}(t)\rmd B_i^\dagger(t)=\delta_{li}\rmd B_k^\dagger(t), \qquad &\rmd \Lambda_{kl}(t)\rmd \Lambda_{ij}(t)=\delta_{li}\rmd \Lambda_{kj}(t);
\end{split}
\end{equation}
all the other possible products vanish. We shall need also the \emph{generalized Weyl operators} $\mathcal{W}(g;V)$, where $g\in L^2(\Rbb;\Cbb^d)$ and $V$ is a unitary operator on $L^2(\Rbb;\Cbb^d)$; these are unitary operators defined by
\begin{equation}\label{Weyl}
\mathcal{W}(g;V)\, e(f) = \exp\left\{\rmi\IM \langle Vf|g\rangle \right\} e(Vf+g), \qquad \forall f\in L^2(\Rbb;\Cbb^d).
\end{equation}
From the definition one obtains the composition law
\begin{equation}\label{Weylprod}
\mathcal{W}(h;V)\,\mathcal{W}(g;U) =  \exp\left\{-\rmi \IM \langle h|Vg\rangle \right\}
\mathcal{W}(h+Vg;VU).
\end{equation}
In the case $V=\openone$, it is possible to show that
\[
\Wcal(g;\openone)=\exp\biggl\{\sum_{k=1}^d\biggl(\int_{-\infty}^{+\infty}g_k(t)\rmd B_k^\dagger(t)- \text{h.c.}\biggr)\biggr\},
\]
where h.c.\ means hermitian conjugate, and from \eqref{Weyl} one sees that $\Wcal(g;\openone)$ is the field analog of what is called a displacement operator in quantum optics \cite{BarG13}.

Let us introduce now a quantum system with separable Hilbert space $\Hscr$; let $H$, $R_k$, $S_{kl}$ be system operators with $H$ self-adjoint and the operator matrix $(S_{kl})$ defining a unitary operator $S$ on $\Hscr\otimes \Cbb^d$. Then, we consider the system/field evolution equation given by the HP-equation
\begin{multline}\label{HPequ}
{\rmd } U(t) = \biggl\{ \sum_k R_k \,{\rmd } B_k^\dagger(t) + \sum_{kl} \left(S_{kl}-
\delta_{kl}\right) {\rmd } \Lambda_{kl}(t)
\\ {}- \sum_{kl} R_k^\dagger S_{kl }\,{\rmd } B_l(t) -
\biggl(\frac 1 2 \sum_{k} R_k^\dagger R_k+\rmi H\biggr){\rmd } t \biggr\}  U(t),
\end{multline}
with the initial condition $U(0) =\openone$. When $H$ and $R_k$ are bounded operators there is a unique solution, which is unitary and strongly continuous in $t$ \cite{Parthas92}. When these operators are unbounded some restrictions are needed in order to control the domains; then, existence, uniqueness, unitarity can be proved \cite{FagW03,Fagnola99}. The solution $U(t)$ gives the evolution in the interaction picture with respect to the free evolution of the field, which is modelled by the so called left time shift $\Theta(t)$ in the Fock space; indeed, $\hat U(t)=\Theta(t)U(t)$, $t\geq 0$, and $\hat U(t)=U(-t)^\dagger \Theta(-t)^\dagger$, $t<0$, defines a strongly continuous unitary group, whose Hamiltonian has been characterized in \cite{Greg01}.

If we now consider a generic system operator $X$, its evolution in the Heisenberg description is given by $X(t)=U(t)^\dagger XU(t)$. By differentiating this product according to the rules of quantum stochastic calculus, summarized by \eqref{Itotab}, and taking into account that $U(t)$ is a unitary operator,
we get the \emph{quantum Langevin equations}
\begin{multline}\label{jdot}
\rmd X(t) = \biggl(\rmi [H(t),X(t)] -\frac 1 2 \sum_k \bigl( R_k(t)^\dagger[R_k(t),X(t)] +
[X(t),R_k(t)^\dagger]R_k(t)\bigr)\biggr) \rmd t \\ {}+ \sum_{kl}S_{lk}(t)^\dagger[X(t),R_l(t)]
\rmd B_k^\dagger(t)
-\sum_{kl}[X(t),R_l(t)^\dagger]S_{lk}(t)\rmd B_k(t) \\ {}+ \sum_{kl}\biggl(\sum_jS_{jk}(t)^\dagger X(t)S_{jl}(t)-\delta_{kl}\biggr)
\rmd \Lambda_{kl}(t).
\end{multline}
If $\rho_0$ is a generic statistical operator for the system and $\sigma$ a field state, we can consider the reduced state of the system $\rho(t)=\Tr_\Gamma\left\{U(t)\rho_0\otimes \sigma U(t)^\dagger\right\}$. When $\sigma$ is the vacuum state or, more generally, a coherent vector, then the reduced system state $\rho(t)$ satisfies a Markovian master equation \cite{Parthas92,Bar06} with a Lindblad type generator \cite{Lin76}. If a more general state is taken for $\sigma$, non-Markov effects enter into play and a simple closed evolution equation for the reduced dynamics could even not exist \cite{BarPer02,Bar06,GJN12}.

Also the fields in the Heisenberg picture can be introduced \cite{GarC85}; these are the \emph{output fields}
\begin{equation}\begin{split}
B_k^{\rm out}(t)=&U(t)^\dagger B_k(t)U(t), \qquad B_k^{\rm out\;\dagger}(t)=U(t)^\dagger B_k^\dagger(t)U(t),\\ &\Lambda_{kl}^{\rm out}(t)=U(t)^\dagger \Lambda_{kl}(t)U(t).
\end{split}
\end{equation}
The outputs fields represent the fields after the interaction with the system, while $B_k(t)$, $B_k^\dagger(t)$, $\Lambda_{kl}(t)$ are the fields before the interaction and, so, they are called \emph{input fields}. By differentiating the products defining the output fields and using \eqref{HPequ} and \eqref{Itotab}, we get the input/output relations \cite{Bar06}
\begin{equation}\label{Bout}
\rmd B_k^{\rm out}(t)=\sum_l S_{kl}(t)\rmd B_k(t)+ R_k(t)\rmd t,
\end{equation}
\begin{multline}\label{Lout}
\rmd \Lambda_{kl}^{\rm out}(t)=\sum_{ij} S_{ki}(t)^\dagger S_{lj}(t)\rmd \Lambda_{ij}(t)+\sum_iS_{ki}(t)^\dagger R_l(t)\rmd B_i^\dagger(t) \\ {} \sum_i+ R_k(t)^\dagger S_{li}(t) \rmd B_i(t)+ R_k(t)^\dagger R_l(t)\rmd t.
\end{multline}
By the properties of $U(t)$ we get $U(T)^\dagger B_k(t)U(T)=U(t)^\dagger B_k(t)U(t)$, $\forall T\geq t$, and similar equations for the other fields. This implies that the output fields satisfy the same CCRs as the input fields.
Self-adjoint combinations of the output fields commuting for different times represent field observables which can be measured with continuity in time and this is the key ingredient for a quantum theory of measurements in continuous time \cite{Bar06,BarCS11,Bar86}.

\section{Langevin \ equations \ for \ a \ mechanical \ oscillator \ in \ a \ thermal \ bath}\label{LangTH}
In this section we present the description of a quantum dissipative mechanical oscillator obtained in \cite[Sects.\ 2, 3]{BarV15}. The Hilbert space of the system is $\Hscr=L^2(\Rbb)$ and $q$ and $p$ denote the usual position and momentum operator in dimensionless units, satisfying the commutation relations [$q,p]=\rmi$. We denote by $\Om>0$ the bare frequency of the mechanical oscillator and by $\mgam>0$ its damping rate; we consider only the underdamped case: $\Om>\mgam/2$. Then, we introduce the damped frequency $\om$ and the phase factor $\tau$ by
\begin{equation}\label{om}
\om=\sqrt{\Omq-\frac{\mgamq }4},
\qquad \tau=\frac{\om}{\Om}-\frac{\rmi}{2} \frac{\mgam }{\Om}.
\end{equation}
We define now the mode operator
\begin{equation}\label{aqptau}
a_\rmm   = \sqrt{\frac{\Om}{2 \om}} \left( q +  {\rmi \tau} p\right)= \frac{1}{\sqrt{2 \om\Om}} \left(\Om  q +  \frac \mgam 2\, p +\rmi \om  p\right),
\end{equation}
satisfying the commutation rules $[a_\rmm,a_\rmm^\dagger]=1$. The inverse transformation turns out to be
\begin{equation}\label{atoqp}
q=\sqrt{\frac\Om{2\om}}\left(\overline \tau \, a_\rmm  +  \tau a_\rmm ^\dagger\right), \qquad p=\rmi\sqrt{\frac{ \Om}{2\om}}\left(a_\rmm ^\dagger - a_\rmm \right).
\end{equation}
We need also the self-adjoint operator
\begin{equation}\label{Hm}
H_\rmm=\frac{\hbar\Om}2\left(p^2+q^2\right) + \frac {\hbar\mgam }  4\left\{q,p\right\} =\hbar \om \left(a_\rmm ^\dagger a_\rmm  +\frac 1 2 \right).
\end{equation}

We introduce now the HP-equation for a mechanical oscillator in a thermal bath by taking in \eqref{HPequ} a single field $B_1(t)\equiv B_{\rm th}(t)$ and  $H=H_\rmm$, $R_1=\sqrt \mgam\, a_\rmm$, $S=\openone$. By \eqref{jdot} the quantum Langevin equations for $a_\rmm$, $q$, $p$ turn out to be
\begin{equation}\label{adot}
\rmd a_\rmm (t)=-\left(\rmi \om+\frac \mgam  2\right) a_\rmm (t)\rmd t - \sqrt{\mgam } \,\rmd B_{\rm th}(t),
\end{equation}
\begin{subequations}\label{qpdot}
\begin{equation}\label{qdot}
\rmd q(t)=\Om p(t) \rmd t +\rmd C_q(t),
\end{equation}
\begin{equation}\label{pdot}
\rmd p(t)=-\bigl(\Om  q(t) +\mgam  p(t)\bigr) \rmd t+ \rmd C_p(t),
\end{equation}
\end{subequations}
in which we have introduced the Hermitian quantum noises
\begin{equation}\label{C_qp}\begin{split}
C_q(t)&= -\sqrt{\frac{\mgam\Om }{2\om}} \left(\overline\tau\, B_{\rm th}(t)+  \tau B_{\rm th}^\dagger(t)\right),
\\
C_p(t)&=\rmi\sqrt{\frac{\mgam\Om }{2\om}}\left(B_{\rm th}(t)- B_{\rm th}^\dagger(t)\right).\end{split}
\end{equation}
By \eqref{intCCR} the new noises
obey the commutation rules
\begin{equation}\label{[CC]}
\left[ C_q(t), C_p(s)\right]=  \rmi\mgam \, {\min\{t,s\}}, \qquad\left[ C_q(t), C_q(s)\right]=\left[ C_p(t), C_p(s)\right]=0.
\end{equation}
Obviously, from \eqref{aqptau}, \eqref{atoqp} we have that the equation \eqref{adot} for $a_\rmm$ is equivalent to the system \eqref{qpdot} for $q$ and $p$. By construction, due to the unitarity of $U(t)$, the commutation relations for the system operators are preserved; also a direct verification is possible by showing that the quantum stochastic differential of $[q(t),p(t)]$ vanishes due to \eqref{[CC]}. Our choice of the field state will be such that the mean values of the noises $C_q(t)$, $C_p(t)$ are vanishing and this gives that the evolution equations for the mean values of $q$ and $p$ coming from \eqref{qpdot} are exactly the classical equations for an underdamped oscillator. This fact is a first justification of the choice \eqref{Hm} for the Hamiltonian and of the not usual connection \eqref{atoqp} of position and momentum with the mode operator.

\subsection{The field state}\label{fieldstate}
As field state we take the mixture of coherent states
\begin{equation}
\sigma_{\rm th}^T=\Ebb[|e(f_T)\rangle\langle e(f_T)|], \qquad f_T(s)=1_{(0,T)}(s)f(s),
\end{equation}
where $f$ is a complex stochastic process with locally square integrable trajectories and $\Ebb$ denotes the expectation with respect to the probability law of the process $f$. In the argument of a coherent vector only square integrable functions are allowed, while the trajectories of the process $f$ are only locally square integrable. So, we have introduced the cutoff $T$,  representing a large time, which we will let tend to infinity in the final formulae describing the quantities of direct physical interest. As explained in \cite[Sect.\ 3.2.1]{BarV15} this is a field analog of the regular $P$-representation for the case of discrete modes \cite{GarZ00}. In quantum optics, mixtures of coherent vectors with respect to true probabilities are interpreted as classical states.

To represent the phonon bath \cite{BarV15} we take $f$ to be a complex Gaussian stationary stochastic process with
vanishing mean, $\Ebb[f(t)]=0$, and correlation functions
\begin{equation}\label{<f>}
\Ebb[f(t)\,f(s)]=0, \qquad \Ebb[\overline{f(t)}\,f(s)]=:F(t-s).
\end{equation}
Thanks to stationarity, the function $F(t)$ is positive definite, so
that according to Bochner's theorem its Fourier transform
\begin{equation}   \label{Nu}
N(\nu)=\int_{-\infty}^{+\infty}\rme^{-\rmi\nu t}F(t)\,\rmd t
\end{equation}
is a positive function, which we assume to be absolutely integrable,
thus implying a finite power spectral density for the process.

By this choice of the state we get that the noises \eqref{C_qp} have vanishing means and symmetrized quantum correlations given by
\begin{multline}
\frac{\partial^2 \ }{\partial t\partial s}\langle\left\{
C_q(t),C_q(s)\right\}  \rangle = \frac{\partial^2 \
}{\partial t\partial s}\langle \left\{  C_p(t),C_p(s)\right\}  \rangle   \\ {}=\mgam\,\frac{\Om}{\om}\left[\delta(t-s) +2\RE F(t-s)\right],\label{qq,pp}
\end{multline}
\begin{equation}
\frac{\partial^2 \ }{\partial t\partial
  s}\langle \left\{  C_q(t),C_p(s)\right\}  \rangle
=2\mgam
 \,\IM F(t-s)-\frac{\mgamq}{2\om}\left[\delta(t-s) +2\RE F(t-s)\right],\label{+qp}
\end{equation}
where $\langle\left\{
C_i(t),C_j(s)\right\}  \rangle := \lim_{T\to+\infty}\Tr_\Gamma\left(\left\{
C_i(t),C_j(s)\right\} \sigma_{\rm th}^T\right)$.
Let us stress that the noises appearing in system \eqref{qpdot} cannot be arbitrary. They have to guarantee the preservation of the commutation relations $[q(t),p(t)]=\rmi$ by satisfying suitable commutations relations, equations \eqref{[CC]} in our case. Moreover, their symmetrized correlations must be compatible with their commutators, as it must be
\[
\sum_{i,j=q,p}\int_0^T\rmd t \int_0^T\rmd s \,\overline{h_i(t)}\,h_j(s)\,\frac{\partial^2 \ }{\partial t\partial s}\langle\left\{
C_i(t),C_j(s)\right\} +[C_i(t),C_j(s) ]\rangle \geq 0, 
\]
$\forall T>0$, for all choices of the test functions $h_i(t)$. Also this property is true in our case because our noises and their correlations are an exact consequence of a unitary model and of the choice of a well defined state. This is not true in other proposals, where the positivity property above is not satisfied or divergences are introduced by not well defined approximation; see the discussion in \cite[Sect.\ 3.3]{BarV15}.

\subsection{The reduced state of the mechanical oscillator}

Let $\rho_0$ be the initial state of the oscillator. It is easy to see that the random reduced state $\Tr_\Gamma\{U(t)\rho_0\otimes |e(f_T)\rangle\langle e(f_T)|U(t)^\dagger\}$ satisfies an usual quantum master equation with random coefficients. But this is not true for its mean, the reduced state
\[
\rho(t)=\Tr_\Gamma\{U(t)\rho_0\otimes \sigma_{\rm th}^T U(t)^\dagger\}, \qquad 0\leq t<T.
\]
By the properties of HP-equation there is no dependence on $T$ as long as $T\geq t$. In any case it is possible to characterize the equilibrium state of the system by solving the linear quantum Langevin equations \eqref{qpdot} and computing the first two moments of $q(t)$ and $p(t)$ for $t\to +\infty$. The reduced  equilibrium state
\[
\rho_{\rm eq}=\lim_{t\to+\infty}\lim_{T\to +\infty}\Tr_\Gamma \left\{U(t) \left(\rho(0)\otimes \sigma_{\rm th}^T
  \right)U(t)^\dagger\right\}
\]
turns out \cite{BarV15} to be a Gaussian state with $\langle q\rangle_{\rm eq}=\langle p\rangle_{\rm eq}=0$ and
\begin{equation}
\langle q^2\rangle_{\rm eq}= \langle p^2\rangle_{\rm eq}=\frac\Om\om \left(N_{\rm eff}+\frac1 2 \right),
\qquad
\langle\left\{ q,p\right\}\rangle_{\rm eq}= -\frac\mgam{\om} \left(N_{\rm eff}+\frac1 2 \right),
\end{equation}
where
\begin{equation}\label{Neff}
N_{\rm eff}:= \frac \mgam{2\pi}\int_\Rbb \frac{N(\nu)}{\frac{\mgamq}4+\left(\om-\nu\right)^2}\,\rmd \nu.
\end{equation}
An important property of our model is that the energy equipartition in mean holds: \\ $\frac{\hbar\Om}2\,\langle q^2\rangle_{\rm eq}= \frac{\hbar\Om}2\,\langle p^2\rangle_{\rm eq}$. Moreover, the mean of the Hamiltonian \eqref{Hm} turns out to be
\[
\langle H_\rmm\rangle_{\rm eq}= \hbar\om\left(N_{\rm eff}+\frac12\right).
\]

\section{Radiation pressure interaction}\label{radpress}

We consider now the case of a mirror mounted on a vibrating structure and directly illuminated by a laser, so that it is subject to a radiation pressure force. One has to add a further interaction term into the HP-equation suitable to produce a force proportional to the photon flux in equation \eqref{pdot} for $p$. If we consider a well collimated laser beam and a perfect mirror, it is possible to represent the light by a single ray impinging on the mirror and reflected according to the laws of the geometrical optics. So, we take $d=2$ and $B_1(t)\equiv B_{\rm th}(t)$, $H=H_\rmm$, $R_1=\sqrt\mgam\, a_\rmm$, $S_{11}=1$ as before; moreover, we add a further field $B_2(t)\equiv B_{\rm em}(t)$, representing the electromagnetic field, and we write $\Lambda_{\rm em}(t)=\Lambda_{22}(t)$. A force proportional to the rate of photon arrivals means to have a term $v\rmd  \Lambda_{\rm em}(t)$ in \eqref{pdot}. By comparing this expression with \eqref{jdot} with $X=p$, one sees that we need $S_{22}\equiv S=\rme^{\rmi\phi}\rme^{\rmi v q}$ and $S_{12}=S_{21}=0$, $R_2=0$; $\phi$ is a phase shift introduced by the mirror.
So, the final HP-equation is
\begin{multline}
\rmd U(t)=\biggl\{-\frac\rmi \hbar\,H_\rmm\rmd t +\sqrt \mgam  \left ( a_\rmm\rmd B_{\rm th}^\dagger(t) -a_\rmm^\dagger\rmd B_{\rm th}(t)\right)
\\
{}+\left(S-1\right)\rmd  \Lambda_{\rm em}(t)
-\frac \mgam {2 }\,a_\rmm^\dagger
a_\rmm\rmd t \biggr\}U(t), \label{eq:U}
\end{multline}
\[
S=\rme^{\rmi\phi}\rme^{\rmi v q}, \quad v\in\Rbb, \quad \phi\in [0,2\pi), \qquad U(0)=\openone.
\]
From \eqref{jdot} one gets the relevant quantum Langevin equations
\begin{equation}\label{eq:a}
\rmd a_\rmm (t)=-\left(\rmi \om+\frac \mgam  2\right) a_\rmm (t)\rmd t - \sqrt{\mgam } \,\rmd B_{\rm th}(t)+\rmi \tau v\sqrt{\frac \Om{2\om}}\,\rmd \Lambda_{\rm em}(t),
\end{equation}
or, equivalently,
\begin{subequations}
\begin{gather}\label{eq:q}
\rmd q(t)=\Om p(t) \rmd t +\rmd C_q(t),
\\  \label{eq:p}
\rmd p(t)=-\left(\Om  q(t) +\mgam  p(t)\right) \rmd t+ \rmd C_p(t)+v \rmd \Lambda_{\rm em}(t);
\end{gather}
\end{subequations}
The quantum noises $C_q(t)$ and $C_p(t)$ are given by \eqref{C_qp}. The linearity of such equations allows for an explicit solution
\begin{multline}\label{a(t)}
a_\rmm (t)=\rme^{-\left(\rmi \om +\frac \mgam  2 \right)t}a_\rmm -\sqrt \mgam  \int_0^t\rme^{-\left(\rmi \om +\frac \mgam  2 \right)\left(t-s\right)}\rmd B_{\rm th}(s) \\ {}+\rmi \tau v\sqrt{\frac \Om{2\om}}\int_0^t\rme^{-\left(\rmi \om +\frac \mgam  2 \right)\left(t-s\right)} \rmd \Lambda_{\rm em}(s) ,
\end{multline}
leading for the position and momentum Heisenberg operators to
\begin{multline}
q(t) =\rme^{-\mgam  t/2} \left(q\cos \om t +\frac {\mgam  q+2\Om p}{2\om} \,\sin\om t\right)
\\ {}-\sqrt{\frac{\Om \mgam  }{2\om}}\biggl\{\overline \tau \int_0^t\rme^{-\left(\rmi \om +\frac \mgam  2\right)(t-s)}\rmd B_{\rm th}(s) +\text{h.c.}\biggr\}
\\ {}+
\frac{\Om v}\om\int_0^t \rme^{-\frac\mgam 2\left(t-s\right)}\sin \om\left(t-s\right)\,\rmd \Lambda_{\rm em}(s),\label{q(t)}
\end{multline}
\begin{multline}
p(t) =\rme^{-\mgam  t/2} \left(p\cos \om t-\frac {2\Om q+\mgam  p}{2\om}\,\sin\om t\right)
\\ {}
+\sqrt{\frac{\Om \mgam  }{2\om}}\biggl\{\rmi\int_0^t\rme^{-\left(\rmi \om +\frac \mgam  2\right)(t-s)}\rmd B_{\rm th}(s) +\text{h.c.}\biggr\}
\\ {}+
v\int_0^t \rme^{-\frac\mgam 2\left(t-s\right)}\left(\cos\om\left(t-s\right)-\frac{\mgam}{2\om}\,\sin \om\left(t-s\right)\right)\rmd \Lambda_{\rm em}(s).\label{p(t)}
\end{multline}

\subsection{Input-output relations}
We now consider the Heisenberg picture for the electromagnetic component of the field:
\begin{equation}
  \label{eq:2}
B_{\rm em}^{\mathrm{out}}(t)=U(t)^\dagger B_{\rm em}(t)U(t),\qquad \Lambda^{\mathrm{out}}_{\rm em}(t)=U(t)^\dagger \Lambda_{\rm em}(t)U(t).
\end{equation}
By \eqref{Bout}, \eqref{Lout} we get the input-output relations
\begin{gather}\label{Beminout}
\rmd  B^{\mathrm{out}}_{\rm em}(t)=S(t)\rmd B_{\rm em}(t)=\rme^{\rmi vq(t)+\rmi\phi}\rmd B_{\rm em}(t),
\\ \label{2Ls}
\rmd \Lambda^{\mathrm{out}}_{\rm em}(t)=S(t)^\dagger S(t) \rmd \Lambda_{\rm em}(t)
=\rmd \Lambda_{\rm em}(t).
\end{gather}
Note that the number operator for the photons is not changed by the interaction with the mirror.

By using \eqref{q(t)} the scattering operator can be decomposed as the product
\begin{equation}\label{SWW}
S(t)=\rme^{\rmi v q(t)+\rmi \phi}=S_0(t)\Wcal_{\rm th}(\ell_t;\openone)\Wcal_{\rm em}(0;V(t)),
\end{equation}
where a system operator and two Weyl operators appear:
\begin{equation}
S_0(t)=\rme^{\rmi \phi}\exp\left\{\rmi v \rme^{-\mgam  t/2} \left(q\cos \om t +\frac {\mgam  q+2\Om p}{2\om} \,\sin\om t\right)\right\}\overset{t\to +\infty}{\longrightarrow} \rme^{\rmi \phi},
\end{equation}
\begin{equation}\label{Wth}
\Wcal_{\rm th}(\ell_t;\openone)=\exp\left\{\int_0^{+\infty} \ell_t(s)\rmd B_{\rm th}^\dagger(s) -\text{h.c.}\right\},
\end{equation}
\begin{equation}\label{Wem}
\Wcal_{\rm em}(0;V(t))=\exp\left\{\rmi
\frac{\Om v^2}\om\int_0^t \rme^{-\frac\mgam 2\left(t-s\right)}\sin \om\left(t-s\right)\,\rmd \Lambda_{\rm em}(s)\right\},
\end{equation}
with
\begin{equation}\label{ell(t)}
\ell_t(\bullet)=-\rmi v\tau \sqrt{\frac{\Om \mgam  }{2\om}}\,1_{(0,t)}(\bullet)\rme^{\left(\rmi \om -\frac \mgam  2\right)(t-\bullet)},
\end{equation}
\begin{equation}\label{V(t)}\begin{split}
\bigl(V(t)u\bigr)(s)=V(s;t)u(s),\qquad &\forall u\in L^2(\Rbb),
\\
V(s;t)=\exp\left\{\rmi v^2 h(t-s)1_{(0,t)}(s)\right\},\qquad &h(r)= \frac\Om\om\,  \rme^{-\frac\mgam2\,r}\sin \om r.
\end{split}
\end{equation}
The Weyl operator $\Wcal_{\rm th}(\ell_t;\openone)$ \eqref{Wth} is a displacement operator with function $\ell_t$ \eqref{ell(t)} acting on the thermal component and $\Wcal_{\rm em}(0;V(t))$ \eqref{Wem} is a Weyl operator acting only on the electromagnetic component and characterized by the unitary operator $V(t)$ \eqref{V(t)}.

\subsection{The field state}\label{sec:fieldstate}
Now the environment is described by a two-component field and its state must describe the phonon bath and the laser light. As field state we take the mixture of coherent states
\begin{equation}
\sigma_{\rm env}^T=\Ebb[|e(u_T)\rangle\langle e(u_T)|], \qquad u_T(s)=1_{(0,T)}(s)u(s),\qquad u(s)=\begin{pmatrix} f(s)\\ g(s)\end{pmatrix},
\end{equation}
where $f$ is the stochastic process described in Section \ref{fieldstate} and $g$ describes a phase-diffusion model of a laser \cite{BarPer02}, namely
\[
g(t)= \lambda\rme^{-\rmi \left(\omega_0 t+ \sqrt{ L_p}\, W(t)\right)}, \qquad \lambda \in \Cbb, \quad \omega_0>0, \quad L_p> 0;
\]
$W(t)$ is a standard Wiener process independent from the process $f$. It is easy to see that
\[
\lim_{T\to +\infty}\biggl[\abs{\frac 1 {\sqrt T}\int_0^T \rme^{\rmi \mu t}g(t)\rmd t}^2\biggr]=\frac {\abs\lambda^2 L_p}{\frac{L_p^{\,2}}4 + \left(\mu-\omega_0\right)^2}\,;
\]
so, the laser light has carrier frequency $\omega_0$ and Lorentzian spectrum of width $L_p$.
A possible generalization would be to take $\lambda \to \lambda(t)$, with $\lambda(t)$ a further stochastic process; this would allow to describe also amplitude fluctuations.

With this choice of the state, for the thermal noises $C_q$ and $C_p$ we have vanishing means and symmetrized correlations \eqref{qq,pp}, \eqref{+qp}, while for the electromagnetic field we get
\begin{equation}
\Tr\left\{\rmd B_{\rm em}(t) \sigma_{\rm env}^T\right\}= \lambda \rme^{-\left(\rmi\omega_0+\frac {L_p}2\right)t}\rmd t,\qquad \Tr\left\{ \rmd \Lambda_{\rm em}(t)\sigma_{\rm env}^T\right\}=\abs\lambda^2\rmd t,
\end{equation}
\begin{equation}\label{<BBem>}
\Tr\left\{\rmd B_{\rm em}^\dagger(s) \rmd B_{\rm em}(t)\sigma_{\rm env}^T\right\}= \abs\lambda^2 \rme^{-\rmi\omega_0\left(t-s\right)-\frac {L_p}2\abs{t-s}}\rmd t\,\rmd s,
\end{equation}
\begin{equation}
\Tr\left\{\rmd \Lambda_{\rm em}(s)\,\rmd \Lambda_{\rm em}(t)  \sigma_{\rm env}^T\right\} =  \left[\delta(t-s)+\abs\lambda^2\right]\abs\lambda^2\rmd t \rmd s.
\end{equation}

\subsection{The equilibrium state of the mechanical oscillator}
\label{sec:physinter}
Again, we can introduce the reduced state of the mechanical oscillator
\[
\rho(t)=\Tr_\Gamma\{U(t)\rho_0\otimes \sigma_{\rm env}^T U(t)^\dagger\}, \qquad 0\leq t<T,
\]
and the reduced  equilibrium state
\[
\rho_{\rm eq}=\lim_{t\to+\infty}\lim_{T\to +\infty}\Tr_\Gamma \left\{U(t) \left(\rho(0)\otimes \sigma_{\rm env}^T
  \right)U(t)^\dagger\right\}.
\]
By working in the Heisenberg picture, from \eqref{a(t)}-\eqref{p(t)} and the moments of the fields we get easily
\begin{gather}
\langle p\rangle_{\rm eq}=0, \qquad \langle q\rangle_{\rm eq}=\frac{v\abs\lambda^2}\Om=:q_\infty\,, \qquad \langle\left\{ q,p\right\}\rangle_{\rm eq}= -\frac\mgam{\om} \left(N_{\rm eff}+\frac1 2 \right),
\\
\langle q^2\rangle_{\rm eq}-q_\infty^2= \langle p^2\rangle_{\rm eq}=\frac\Om\om \left(N_{\rm eff}+\frac1 2 \right)+\frac{ \abs\lambda^2 v^2}{2\mgam},\label{q2p2}
\end{gather}
where $ N_{\rm eff}$ is given by \eqref{Neff}. By \eqref{q2p2} the energy equipartition in mean holds again for the fluctuation part.
Moreover, the mechanical mode occupancy is given by
\[
\langle a_\rmm^\dagger a_\rmm\rangle_{\rm eq} -\frac { \Om q_\infty^2}{2\om}=  N_{\rm eff} + \frac { \Om v^2\abs\lambda^2}{2\om\mgam};
\]
we have also
\[
\langle a_\rmm^{\,2}\rangle_{\rm eq}-\frac {\Om q_\infty^2}{2\om}= \frac{ \abs\lambda^2 v^2}{4\om\Om}\left(
\frac\mgam 2 +\rmi \om\right)=\rmi \tau \, \frac{ \abs\lambda^2 v^2}{4\om}\,.
\]

Finally, it is possible to show that in the limiting case of constant phonon spectrum, $N(\nu)\to N_{\rm eff}$, and no phase diffusion, $L_p\downarrow 0$, the reduced system state satisfies a Markovian master equation with  Liouville operator
\begin{multline*}
\Lcal[\rho]=-\frac{\rmi}{\hbar} \left[ H_\rmm ,\rho\right]+\mgam  \left(N_{\rm eff}+1\right) \left( a_\rmm \rho a_\rmm ^\dagger
- \frac 12 \left\{a_\rmm ^\dagger a_\rmm ,\rho\right\}\right)
\\ {}+\mgam  N_{\rm eff}\left( a_\rmm^\dagger \rho a_\rmm
- \frac 12 \left\{a_\rmm a_\rmm ^\dagger ,\rho\right\}\right)
+\abs\lambda^2\left(\rme^{\rmi v q}\rho\rme^{-\rmi v q}-\rho\right).
\end{multline*}
The last term is new and describes the momentum kicks due to the scattering of photons. The other terms of the Liouville operator have the appearance of an usual generator for the dynamics of a mode in a thermal bath; however, the important point is that the link of the mode operator with position and momentum is not the usual one, but it is given by \eqref{aqptau}, \eqref{atoqp} \cite[Sect.\ 2.2]{BarV15}.

\section{Heterodyne detection}\label{hetdet}

To get information on the mechanical oscillator we can detect in various ways and analyse the light reflected by the vibrating mirror. In the balanced heterodyne detection scheme the light coming from our system is made to beat with a strong laser field (the local oscillator); the
light impinging on the mirror and the local oscillator are produced by different laser sources; the stimulating laser
frequency $\omega_0$ and the local oscillator frequency, say $\mu$, are in
general different. Moreover, the phase difference cannot be maintained stable
and this erases some interference terms. It can be shown
\cite[Sect.\ 3.5]{Bar06} that the balanced heterodyne detection scheme
corresponds to the measurement in continuous time of the observables
\begin{equation}\label{def:I}
I(\mu;t)=\int_0^t \sqrt \varkappa \,\rme^{-\varkappa \left(t-s\right)/2}\,
\rme^{\rmi\mu s+\rmi\alpha}\,\rmd B_{\rm em} (s)+\text{h.c.},
\end{equation}
where $\alpha$ is a phase depending on the optical paths and $\sqrt \varkappa
\,\rme^{-\varkappa t/2}$, \ $\varkappa>0$, represents the detector response
function. In the Heisenberg description the observables become the ``output current''
\begin{equation}\label{def:Iout}
I_{\rm out}(\mu;t)=U(t)^\dagger I(\mu;t)U(t).
\end{equation}
By using \eqref{Beminout} we obtain the explicit expression
\begin{equation}\label{IJ}
I_{\rm out}(\mu;t)=J(t)+{\rm h.c.},
\qquad
J(t)=\sqrt\varkappa \,\rme^{\rmi \left(\alpha+\phi\right) }\int_0^t\rme^{-\frac\varkappa 2 \left(t-s\right)+\rmi \mu s}\rme^{\rmi vq(s)}\rmd B_{\rm em}(s).
\end{equation}
By the definition of $I(\mu;t)$ and the properties of $U(t)$ (see the discussion at the end of Section \ref{intro}) we get $[I(\mu;t),I(\mu;s)]=[I_{\rm
out}(\mu;t),I_{\rm out}(\mu;s)]=0$, which says that the output current at time
$t$ and the current at time $s$ are compatible observables. Note that to change $\mu$ means to change the frequency of the local oscillator, that is to change the measuring apparatus. In general $I_{\rm out}(\mu;t)$ and $I_{\rm out}(\mu';s)$ do not commute, even for $t=s$.

By the rules of quantum mechanics, once one has the commuting observables $I_{\rm out}(\mu;t)$, $t\geq 0$, and the system/field state $\rho_0\otimes\sigma_{\rm env}^T$, the probability law of the stochastic process representing the output of the detection apparatus is obtained \cite{Bar86,Bar06}. By taking the second moment of the output current the mean output power is obtained \cite{Bar06,BarV15} and at large times it turns out to be proportional to
\begin{equation}\label{def:power}
P(\mu)= \lim_{T\to +\infty}\frac 1 T\int_0^T \langle I_{\rm out}(\mu;t)^2\rangle_T\rmd t, \qquad \langle \bullet \rangle_T:= \Tr\left\{\bullet\;\rho_0\otimes\sigma_{\rm env}^T
\right\};
\end{equation}
the limit is in the sense of the distributions in $\mu$. As a function of
$\mu$, $P(\mu)$ is known as \emph{power spectrum}.

By using directly \eqref{def:power}, \eqref{def:Iout}, \eqref{def:I}, \eqref{2Ls}, without computing the explicit expression of $P(\mu)$, one gets easily the ``total output power''
\begin{equation}
\frac1{2\pi}\int_\Rbb \rmd \mu\left[P(\mu)-1\right]= 2\abs\lambda^2\label{totP}.
\end{equation}
For sake of comparison it is interesting to have also the power spectrum of the input light; by setting $\kappa:=\varkappa+L_p$, we get
\begin{equation}\label{Pin}
P_{\rm in}(\mu)= \lim_{T\to+\infty}\frac 1 T \int_0^T\langle I(\mu;t)^2\rangle_T \rmd t=1+\frac {2\abs\lambda^2\kappa }{\frac{\kappa^2}4+\left(\mu-\omega_0\right)^2};
\end{equation}
the final explicit expression in \eqref{Pin} is easily computed by using \eqref{<BBem>} and the CCRs. Moreover, we have immediately
\begin{equation}
\frac1{2\pi}\int_\Rbb \rmd \mu\left[P_{\rm in}(\mu)-1\right]= 2\abs\lambda^2.
\end{equation}
Let us stress that the equality of the total input and output powers is essentially due to \eqref{2Ls}.

\subsection{Exact results}
The explicit expression of the power spectrum can be computed, as we shall show below. Firstly, \eqref{def:power} reduces to
\begin{equation}\label{Pmu1}
P(\mu)=1+2 \lim_{T\to+\infty}\frac 1 T \int_0^T \langle J(t)^\dagger J(t)\rangle_T\rmd t;
\end{equation}
then, we obtain
\begin{multline}\label{Pmu2}
P(\mu)=1 + 4\abs\lambda^2\exp\biggl\{2\abs\lambda^2\RE\int_0^{+\infty}\rmd u  \left(\rme^{\rmi v^2 h(u)}-1\right)-\frac{\left(N_{\rm eff}+\frac 1 2 \right)\Om}{\om}\,v^2\biggr\}\\ {}\times
\RE\int^{+\infty}_0\rmd t\,  \rme^{\left(\rmi \left(\mu-\omega_0\right) -\frac{\kappa}2\right) t}
\exp\biggl\{ \abs{\lambda}^2 \int_0^{+\infty}\rmd s \left[\rme^{\rmi v^2 h(t+s)}-1\right]\left[\rme^{-\rmi v^2h(s)}-1\right]
\\ {}+\int_\Rbb \rmd \nu \, \frac{\Om\mgam v^2\left[ \left(N(\nu)+1\right)\rme^{\rmi\nu t} +N(\nu)\rme^{-\rmi\nu t}\right]}{4\pi\om\left(\frac\mgamq 4 +\left(\nu-\om\right)^2\right)}\biggr\},
\end{multline}
where $h(t)$ is given in \eqref{V(t)} and $\kappa=\varkappa+L_p$. Let us stress that this is an exact result obtained from a unitary quantum evolution and the monitoring in continuous time of commuting observables. Note that in this expression the thermal contributions, the terms containing $N(\nu)$, and the electromagnetic contributions, the terms containing the function $h$, are completely interlaced.

\subsubsection{Proof of equations \eqref{Pmu1} and \eqref{Pmu2}} Let us sketch now the proof of the previous formulae.
By \eqref{IJ} we have
\[
\langle I_{\rm out}(\mu;t)^2\rangle_T=2\RE\langle J(t)^2\rangle_T+ \langle J(t)J(t)^\dagger\rangle_T+ \langle J(t)^\dagger J(t)\rangle_T.
\]
By the presence of the limit in \eqref{def:power}, these terms contribute to $P(\mu)$ only with their large time behaviour. By using \eqref{Beminout}, \eqref{SWW}--\eqref{V(t)}, \eqref{<BBem>} we get
\begin{multline*}
\langle J(t)^2\rangle_T\simeq \varkappa \lambda^2\rme^{2\rmi \left(\alpha+\phi\right)}
\int_0^t\rmd s \int_0^t \rmd r \,\rme^{-\varkappa\left(t-\frac{s+r}2\right)+\rmi \left(\mu-\omega_0\right) \left(s+r\right)-\frac{L_p}2\abs{s-r}-2L_p\left(s\wedge r\right)}
\\ {} \times V(s;r)\langle \Wcal_{\rm th}(\ell_s;\openone) \Wcal_{\rm th}(\ell_r;\openone) \rangle_T \langle \Wcal_{\rm em}\bigl(0;V(s)\bigr) \Wcal_{\rm em}\bigl(0;V(r)\bigr)\rangle_T,
\end{multline*}
\begin{multline}\label{JdagJ}
\langle J(t)^\dagger J(t)\rangle_T\simeq 2
\varkappa\abs\lambda^2\RE\int^t_0\rmd s \int^s_0\rmd r\, \rme^{-\varkappa \left(t-\frac{s+r}2\right)+\rmi \left(\mu-\omega_0\right) \left(s-r\right)-\frac{L_p}2\left(s-r\right)}
\\ {}\times\langle \Wcal_{\rm th}(\ell_r;\openone)^\dagger \Wcal_{\rm th}(\ell_s;\openone) \rangle_T \langle \Wcal_{\rm em}\bigl(0;V(r)\bigr)^\dagger \Wcal_{\rm em}\bigl(0;V(s)\bigr) \rangle_T,
\end{multline}
\begin{multline*}
\langle J(t) J(t)^\dagger\rangle_T-1\simeq 2\varkappa\abs \lambda^2\RE
\int_0^t\rmd s \int_0^s \rmd r \,V(r;s) \rme^{-\varkappa\left(t-\frac{s+r}2\right)+\rmi \left(\mu -\omega_0\right)\left(s-r\right)}
\\ {}\times \rme^{-\frac{L_p}2\left(s-r\right)}\langle \Wcal_{\rm th}(\ell_s;\openone) \Wcal_{\rm th}(\ell_r;\openone)^\dagger \rangle \langle \Wcal_{\rm em}\bigl(0;V(s)\bigr)\Wcal_{\rm em}\bigl(0;V(r)\bigr)^\dagger \rangle_T.
\end{multline*}
Then, one can check that $\lim_{t\to +\infty}\lim_{T\to +\infty}\langle J(t)^2\rangle_T=0$. Moreover, by using the composition law \eqref{Weylprod} for Weyl operators and \eqref{ell(t)}, \eqref{V(t)}, we get
\[
\Wcal_{\rm em}\bigl(0;V(s)\bigr)\Wcal_{\rm em}\bigl(0;V(r)\bigr)^\dagger=\Wcal_{\rm em}\bigl(0;V(r)\bigr)^\dagger\Wcal_{\rm em}\bigl(0;V(s)\bigr)
\]
and, for $s,\,r$ large and $s>r$,
\begin{multline*}
V(r;s)\Wcal_{\rm th}(\ell_s;\openone) \Wcal_{\rm th}(\ell_r;\openone)^\dagger \\ {}=V(r;s)\Wcal_{\rm th}(\ell_r;\openone)^\dagger \Wcal_{\rm th}(\ell_s;\openone) \exp\left\{2\rmi \IM \langle \ell_s|\ell_r\rangle\right\}\simeq \Wcal_{\rm th}(\ell_r;\openone)^\dagger\Wcal_{\rm th}(\ell_s;\openone).
\end{multline*}
This gives $
\langle J(t) J(t)^\dagger\rangle_T\simeq 1+\langle J(t)^\dagger J(t)\rangle_T$ and, so, \eqref{Pmu1} is proved.

Let us consider now \eqref{JdagJ}; recall that $h(u)$ and $V(s,t)$ are given in \eqref{V(t)} and $\ell_t(s)$ in \eqref{ell(t)}. Firstly, the electromagnetic contribution gives, for $s>r$,
\begin{multline*}
\langle \Wcal_{\rm em}\bigl(0;V(r)\bigr)^\dagger \Wcal_{\rm em}\bigl(0;V(s)\bigr) \rangle_T=  \exp\biggl\{\int_0^{s}\Bigl(\overline{V(u;r)}\,V(u;s)
-1\Bigr) \abs{\lambda}^2\rmd u\biggr\}
\\ {}=
a(s)\,\overline{a(r)}\,\exp\biggl\{ \abs{\lambda}^2 \int_0^{r}\rmd u \left[\rme^{\rmi v^2 h(s-u)}-1\right]\left[\rme^{-\rmi v^2h(r-u)}-1\right]
\biggr\}
\\ {}\simeq
\abs{a(\infty)}^2\exp\biggl\{ \abs{\lambda}^2 \int_0^{r}\rmd u \left[\rme^{\rmi v^2 h(s-u)}-1\right]\left[\rme^{-\rmi v^2h(r-u)}-1\right]
\biggr\},
\end{multline*}
with
\[
a(s)=\exp\biggl\{ \abs\lambda^2\int_0^{s}\rmd u  \left(\rme^{\rmi v^2 h(u)}-1\right)\biggr\}.
\]
Then, again for $s>r$ and both large, the thermal contribution gives
\begin{multline*}
\langle \Wcal_{\rm th}(\ell_r;\openone)^\dagger \Wcal_{\rm th}(\ell_s;\openone) \rangle_T \\ {}=\Ebb\left[\exp\left\{2\rmi \IM \langle f |\ell_s-\ell_r\rangle + \langle \ell_r|\ell_s\rangle-\frac{\norm{\ell_s}^2 +\norm{\ell_r}^2}2\right\}\right]
\\ {}=
\exp\biggl\{-\frac 1 2 \left(\int_0^{s}\abs{\ell_s(u)}^2\rmd u+\int_0^{r}\abs{\ell_r(u)}^2\rmd u\right)+\int_0^{r} \overline {\ell_r(u)}\,\ell_s(u) \rmd u  \\{} -\frac 1 {2\pi }\int_\Rbb \rmd \nu\, N(\nu) \abs{\int_0^{s} \rmd u \,\rme^{\rmi u\nu}\bigl(\ell_s(u)-\ell_r(u)\bigr)}^2\biggr\}
\simeq
\exp\biggl\{-\frac{\left(2N_{\rm eff}+1 \right)\Om v^2}{2\om}
\\ {}+\frac{\Om v^2}{2\om} \rme^{\left(\rmi \om -\frac\mgam 2 \right) \left(s-r\right)}
+\int_\Rbb \rmd \nu \, \frac{\Om\mgam N(\nu)v^2\cos \nu (s-r)}{2\pi\om\left(\frac\mgamq 4 +\left(\om -\nu\right)^2\right)}\biggr\}
\\ {}=
\exp\biggl\{-\frac{\left(2N_{\rm eff}+1 \right)\Om v^2}{2\om}
+\int_\Rbb \rmd \nu \, \frac{\Om\mgam v^2\left[ \left(N(\nu)+1\right)\rme^{\rmi\nu (s-r)}+N(\nu)\rme^{-\rmi\nu (s-r)}\right]}{4\pi\om\left(\frac\mgamq 4 +\left(\om -\nu\right)^2\right)}\biggr\}.
\end{multline*}
By inserting these results into \eqref{JdagJ} and the expression found into \eqref{Pmu1}, we get the final result \eqref{Pmu2}.

\subsection{Linear response}
When the laser light is used as a probe to get information on the dissipative oscillator, the beam can be taken to be weak, which means $\abs\lambda^2$ small. In this case only the linear response is important and we can simplify \eqref{Pmu2} by considering only the ``optical susceptibility''
\[
\Sigma(\mu):=\lim_{\abs\lambda\downarrow 0}\frac {P(\mu)-1}{\abs\lambda^2}.
\]
By \eqref{Pmu2} the weak probe limit gives immediately
\begin{multline}\label{Sigma}
\Sigma(\mu)=
4\exp\biggl\{-\frac{\left(N_{\rm eff}+\frac 1 2 \right)\Om}{\om}\,v^2\biggr\} \RE\int^{+\infty}_0\rmd t\,  \rme^{\left(\rmi \left(\mu-\omega_0\right) -\frac{\kappa}2\right) t}
\\ {}\times \exp\biggl\{\int_\Rbb \rmd \nu \, \frac{\Om\mgam v^2\left[ \left(N(\nu)+1\right)\rme^{\rmi\nu t} +N(\nu)\rme^{-\rmi\nu t}\right]}{4\pi\om\left(\frac\mgamq 4 +\left(\nu-\om \right)^2\right)}\biggr\}.
\end{multline}
Now, the power spectrum is
\begin{equation}
P(\mu)\simeq 1+\abs\lambda^2\Sigma(\mu)
\end{equation}
and we see that in this limit the thermal contribution is completely unaffected by the electromagnetic one. So, we can use the optical probe as a mean to gain information on the mechanical occupancy spectrum $N(\nu)$.

To compute the time integral in \eqref{Sigma} one needs to develop the last exponential in a power series. The result is much more clear when
$N(\nu)$ is slowly varying in a neighbourhood of $\om$ of width $\mgam$. In this case we can made the approximation $N(\nu)\simeq N(\om)$ in the last line of \eqref{Sigma}; by \eqref{Neff} we have also $N_{\rm eff}\simeq N(\om)$. By power expansion we get
\begin{multline}
\Sigma(\mu)\simeq
4\exp\biggl\{-\frac{\left(N(\om)+\frac 1 2 \right)\Om}{\om}\,v^2\biggr\} \RE\int^{+\infty}_0\rmd t\,  \rme^{\left(\rmi \left(\mu-\omega_0\right) -\frac{\kappa}2\right) t}
\\ {}\times \exp\biggl\{\frac{\Om v^2}{2\om}\left[ \left(N(\om)+1\right)\rme^{\left(\rmi\om-\frac\mgam 2 \right) t} +N(\om)\rme^{-\left(\rmi\om+\frac\mgam 2 \right)t}\right]\biggr\}
\\ {}=
2\exp\biggl\{-\frac{\left(N(\om)+\frac 1 2 \right)\Om}{\om}\,v^2\biggr\} \\ {}\times \sum_{m=0}^\infty \sum_{j=0}^m\frac{\Omega_\rmm^{\,m} v^{2m}}{j!(m-j)!2^m\omega_\rmm^{\,m}}\, \frac{\bigl(N(\om)+1\bigr)^jN(\om)^{m-j}\left(\kappa+m\mgam\right)} {\frac{\left(\kappa+m\mgam\right)^2}4+\left[\mu-\omega_0-\left(m-2j\right) \om \right]^2 }.
\end{multline}
So, $\Sigma(\mu)$ appears to be a series of peaks centred on $\omega_0\pm n \om$ and we write
\begin{equation}
\Sigma(\mu)\simeq 2\exp\biggl\{-\frac{\left(N(\om)+\frac 1 2 \right)\Om}{\om}\,v^2\biggr\} \sum_{n\in \Zbb}\Pi_n(\mu).
\end{equation}
By reorganizing the sums we get the expressions of the various peaks and by integration their weights.
\begin{itemize}
\item The peak centred in $\omega_0$:
\begin{equation}
\Pi_0(\mu)=\sum_{j=0}^\infty \frac{\Omega_\rmm^{2j} v^{4j} \bigl(N(\om)+1\bigr)^jN(\om)^j\left(\kappa+2j\mgam\right) }{\left(j!\right)^2 4^j \omega_\rmm^{2j} \left[\frac{\left(\kappa+2j\mgam\right)^2}4+\left(\mu-\omega_0 \right)^2\right]}\,;
\end{equation}
here the term with $j=0$ represents the elastic scattering of photons, while a term with $j>0$ represents the scattering of a photon with exchange with the mechanical oscillator of $j$ energy quanta $\om$. The weight of the peak is
\begin{equation}
\frac 1 {2\pi}\int_\Rbb\Pi_0(\mu)\rmd \mu=\sum_{j=0}^\infty \frac 1 { \left(j!\right)^2} \left(\frac{\Omq v^{4} \bigl(N(\om)+1\bigr)N(\om)}{ 4 \omq }\right)^j\,.
\end{equation}
For $N(\om)=0$ the previous formulae reduce to
\begin{equation}
\Pi_{0}(\mu)= \frac{\kappa }{ \frac{\kappa^2}4+\left(\mu-\omega_0 \right)^2},\qquad \frac 1 {2\pi}\int_\Rbb\Pi_0(\mu)\rmd \mu=1.
\end{equation}
\item The peaks centred in $\omega_0-n\om$, $n=1,2,\ldots$, (Stokes scattering):
\begin{equation}
\Pi_{-n}(\mu)=\sum_{j=0}^\infty \frac{\Omega_\rmm^{2j+n} v^{4j+2n}\bigl(N(\om)+1\bigr)^{j+n}N(\om)^j\bigl(\kappa+\left(2j+n\right)\mgam\bigr) }{j!(j+n)! 2^{2j+n} \omega_\rmm^{2j+n}\left[\frac{\bigl(\kappa+\left(2j+n\right)\mgam\bigr)^2}4+\left(\mu-\omega_0 +n\om \right)^2 \right]}\, ;
\end{equation}
here the term with $j=0$ represents the cession of a quantum $\om$ from the photon to the mechanical oscillator, while a term with $j>0$ represents the same process plus the exchange of other $j$ quanta. The weight is
\begin{equation}
\frac 1 {2\pi}\int_\Rbb\Pi_{-n}(\mu)\rmd \mu=\bigl(N(\om)+1\bigr)^{n}\sum_{j=0}^\infty \frac{\Omega_\rmm^{2j+n} v^{4j+2n}\bigl(N(\om)+1\bigr)^{j}N(\om)^j }{j!(j+n)! 2^{2j+n} \omega_\rmm^{2j+n}}\, .
\end{equation}
For $N(\om)=0$ we get
\begin{gather}
\Pi_{-n}(\mu)= \frac{\Omega_\rmm^{\,n} v^{2n}\bigl(\kappa+n\mgam\bigr) }{n! 2^{n} \omega_\rmm^{\,n} \left[\frac{\bigl(\kappa+n\mgam\bigr)^2}4+\left(\mu-\omega_0 +n\om \right)^2 \right]}\, ,\\ \frac 1 {2\pi}\int_\Rbb\Pi_{-n}(\mu)\rmd \mu= \frac{\Omega_\rmm^{\,n} v^{2n} }{n! 2^{n} \omega_\rmm^{\,n} }\,.
\end{gather}
\item The peaks centred in $\omega_0+n\om$, $n=1,2,\ldots$, (anti-Stokes scattering):
\begin{equation}
\Pi_{n}(\mu)=\sum_{j=0}^\infty \frac{\Omega_\rmm^{2j+n} v^{4j+2n}\bigl(N(\om)+1\bigr)^{j}N(\om)^{j+n}\bigl(\kappa+\left(2j+n\right)\mgam\bigr) }{j!(j+n)! 2^{2j+n} \omega_\rmm^{2j+n}\left[\frac{\bigl(\kappa+\left(2j+n\right)\mgam\bigr)^2}4+\left(\mu-\omega_0 -n\om \right)^2\right]}\, ,
\end{equation}
here the term with $j=0$ represents the cession of a quantum $\om$ from  the mechanical oscillator to the photon, while a term with $j>0$ represents the same process plus the exchange of other $j$ quanta.
Note that the weight turns out to be
\begin{equation}
\frac 1 {2\pi}\int_\Rbb\Pi_{n}(\mu)\rmd \mu=\left(\frac {N(\om)}{N(\om)+1}\right)^n\,\frac 1 {2\pi}\int_\Rbb\Pi_{-n}(\mu)\rmd \mu.
\end{equation}
For $N(\om)=0$ we get $\Pi_{n}(\mu)=0$.
\end{itemize}

The asymmetry between Stokes and anti-Stokes scattering is the base for using the the optical probe as a device for thermometry at low temperatures. Indeed, we have
\begin{equation}
N(\om)=\frac{\int_\Rbb\Pi_{1}(\mu)\rmd \mu}{\int_\Rbb\Pi_{-1}(\mu)\rmd \mu-\int_\Rbb\Pi_{1}(\mu)\rmd \mu}
\end{equation}
and this quantity can be estimated by the area under the curve of the experimental data when the peaks in $\omega_0\pm \om$ are well separated from the elastic peak in $\omega_0$, which means that the widths $\mgam$ and $\kappa=\varkappa+L_p$ are sufficiently small. The \emph{resolved-sideband thermometry} is a technique already used in somewhat similar situations \cite{SN-P13,asp14}.

\end{document}